\documentclass[aps,pra,twocolumn,tightenlines,10pt,showpacs]{revtex4}
\usepackage{amsfonts}
\usepackage{amsmath}
\usepackage{amssymb}
\usepackage{graphicx}
\usepackage{pslatex}

\begin{document}

\preprint{}
\title{Stable giant vortex annuli in microwave-coupled atomic condensates}
\author{Jieli Qin and Guangjiong Dong}
\affiliation{State Key Laboratory of Precision Spectroscopy, Department of Physics, East
China Normal University, Shanghai, China\\
Collaborative Innovation Center of Extreme Optics, Shanxi University,
Taiyuan, Shanxi 030006, People's Republic of China}
\author{Boris A. Malomed}
\affiliation{Department of Physical Electronics, School of Electrical Engineering,
Faculty of Engineering, Tel Aviv University, Ramat Aviv 69978, Israel\\
Laboratory of Nonlinear-Optical Informatics, ITMO University, St. Petersburg
197101, Russia}

\begin{abstract}
Stable self-trapped vortex annuli (VAs) with large values of topological
charge $S$ (\textit{giant VAs}) are not only a subject of fundamental
interest, but are also sought for various applications, such as quantum
information processing and storage. However, in conventional atomic
Bose-Einstein condensates (BECs) VAs with $S>1$ are unstable. Here, we
demonstrate that robust self-trapped fundamental solitons (with $S=0$) and
bright VAs (with the stability checked up to $S=5$), can be created in the
free space by means of the local-field effect (the feedback of the BEC on
the propagation of electromagnetic waves) in a condensate of two-level atoms
coupled by a microwave (MW) field, as well as in a gas of MW-coupled
fermions with spin $1/2$. The fundamental solitons and VAs remain stable in
the presence of an arbitrarily strong repulsive contact interaction (in that
case, the solitons are constructed analytically by means of the Thomas-Fermi
approximation). Under the action of the attractive contact interaction with
strength $\beta $, which, by itself, would lead to collapse, the fundamental
solitons and VAs exist and are stable, respectively, at $\beta <\beta _{\max
}(S)$ and $\beta <\beta _{\mathrm{st}}(S)$, with $\beta _{\mathrm{st}%
}(S=0)=\beta _{\max }(S=0)$ and $\beta _{\mathrm{st}}(S\geq 1)<\beta _{\max
}(S\geq 1)$. Accurate analytical approximations are found for both $\beta _{%
\mathrm{st}}$ and $\beta _{\max }$, with $\beta _{\mathrm{st}}(S)$ growing
linearly with $S$. Thus, higher-order VAs are \emph{more robust} than their
lower-order couterparts, on the contrary to what is known in other systems
that may support stable self-trapped vortices. Conditions for the
experimental realizations of the VAs are discussed.
\end{abstract}

\maketitle

\section{Introduction}

Light and microwaves (MWs) are important tools for controlling dynamics of
atomic Bose-Einstein condensates (BECs). In addition to creating traps and
optical lattices \cite{OL}, various optical patterns, including vortices,
have potential application in the realm of quantum data processing, as the
light patterns can be stored in the form of intrinsic atomic states in BEC,
and released back in the optical form \cite{storage-vortex}. Furthermore,
light can generate entangled vortices in separated condensates \cite{remote}.

The BEC\ feedback on the light propagation, i.e., the \textit{local field
effect} (LFE), may lead to the creation of hybrid light-matter states \cite%
{dong1,dong2,dong3,can,fir,bub}. The electric LFE explains asymmetric
matter-wave diffraction \cite{dong1,li} and predicts polaritonic solitons in
soft optical lattices \cite{dong2}. Further, the magnetic LFE couples MWs to
a pseudo-spinor (two-component) BEC of two-level atoms, thus opening the way
to the creation of hybrid microwave-matter-wave solitons \cite{dong3}. On
the other hand, in current experiments with the pseudo-spinor BECs, atoms
are first transferred to an intermediate level using a MW field, and then
further driven to a target level using a radiation-frequency fields, which
would not allow one to observe manifestations of the magnetic LFE. This
should become possible if the experiments can be performed with the MW field
directly transferring the atoms between the two relevant states.

The LFE plays an increasingly important role in BEC with the increase of the
number of atoms, which can exceed $10^{8}$, as predicted theoretically \cite%
{large} and demonstrated experimentally \cite{large1}, allowing the
LFE-induced long-range interactions between atoms \cite{dong2,dong3} to
produce new manifestations of nonlocal physics. Actually, the long-range
interaction may cover the whole gas, in contrast with fast-decaying nonlocal
interactions in optics \cite{lc} and in dipolar BEC \cite%
{cr,dy,er,polar,dipv,DDreview}. Unlike the species-dependent dipolar forces
\cite{cr,dy,er,polar}, the LFE-induced interaction can be realized in any
ultracold atomic or molecular gas \cite{dong3}.

The LFE was not previously explored in two- and three-dimensional (2D and
3D) settings, where it may give rise to new phenomenology in comparison with
the recently investigated 1D case \cite{dong1,dong2,dong3}, as the
LFE-induced interaction is determined by the underlying Green's function,
which has different forms in effectively 1D, 2D, and 3D geometries (note
that the above-mentioned ``massive" BEC, with a large number
of atoms $\gtrsim 10^{8}$, can be readily morphed into a low-dimensional
shape \cite{large1}). In particular, we demonstrate here that solitary
vortices, alias vortex annuli (VAs), readily self-trap in the 2D setting.
Vortices in BEC are essential for simulating various effects from condensed
matter \cite{bkt}, and as building blocks of quantum turbulence \cite{BY}.
They also help to emulate gravitational physics \cite{gp}, and find
applications, such as phase qubits \cite{pha} and matter-wave Sagnac
interferometers for testing the rotational-equivalence principle \cite%
{you,luo}. As mentioned above, atomic-matter vortices can store and release
information delivered by optical vortex beams \cite{storage-vortex}.

The stabilization of VAs with large values of the topological charge
(vorticity) $S$, which is required for deterministic creation of vortices
\cite{deterministic-creation} and for applications (in particular, the
storage of higher-order optical vortices in the form of their atomic
counterparts), is a challenging issue \cite{fet}. Under repulsive
interactions, vortices supported by a nonzero background are stable solely
for $S=1$, while vortices with $S\geq 2$ split into ones with $S=1$ \cite%
{fet}. For the above-mentioned applications, most relevant are bright VAs in
BEC with attractive nonlinearity. Unlike nonlinear optics, where VAs can be
stabilized by non-Kerr nonlinearities \cite{non-Kerr}, in BEC with
attractive interactions the only setting which gives rise to stable 2D \cite%
{SOC2D} and 3D \cite{SOC3D} \textit{semi-vortices} (with $S_{\uparrow }=1$
and $S_{\downarrow }=0$ in their two components) in the free space is
provided by the spin-orbit coupling. However, all higher-order states, with $%
S_{\uparrow }=1+s$, $S_{\downarrow }=s\geq 1$, are unstable. The family of
single-component modes with $S=1$ may be partly stabilized by a trapping
potential, but all the higher-order VAs with $S\geq 2$ remain unstable in
this case too \cite{trap}. Partly stable VAs with $S\geq 2$ were predicted
only in ``exotic" settings, with the local strength of the
repulsive nonlinearity in the space of dimension $D$ growing with distance $%
r $ from the center faster than $r^{D}$ \cite{Barcelona}, or making use of a
combination of a trapping potential and a \emph{spatially localized}
attractive interaction \cite{Nal}.

In this work, we introduce a 2D hybrid system consisting of a pseudo-spinor
BEC whose two components are coupled by a MW field through a magnetic-dipole
transition. The system gives rise to \emph{stable }giant VAs, i.e., ones with%
\emph{\ arbitrarily high} values of $S$ (the stability checked up to $S=5$).
This is as well possible in the presence of additional contact repulsive
interactions. On the other hand, under the action of an attractive contact
interaction, with strength $\beta $, which drives the critical collapse in
the 2D geometry \cite{collapse}, the VAs exist and are stable, respectively,
for $\beta <\beta _{\max }$ and $\beta <\beta _{\mathrm{st}}\leq \beta
_{\max }$. We demonstrate, by means of analytical and numerical
considerations, that $\beta _{\mathrm{st}}$ linearly grows with $S$, thus
making higher-order vortices \emph{more robust} than lower-order ones,
opposite to what is known in few other models capable to support stable
higher-order VAs \cite{Barcelona,Nal}. It is relevant to mention that the
concept of giant vortices is known in the usual BEC settings with the
contact repulsion \cite{giant}, where they are not self-trapped objects,
i.e., they are not VAs.

The rest of the paper is organized as follows. The model is introduced in
Section II, numerical and analytical results are collected in Section III,
and the paper is concluded by Section IV.

\begin{figure}
\includegraphics{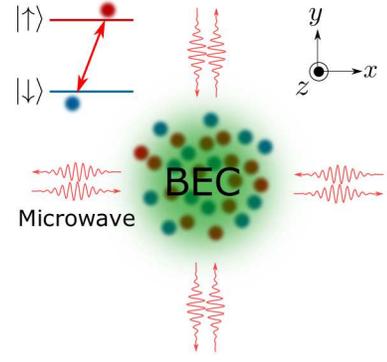}
\caption{Two hyperfine atomic states
coupled by the MW (microwave) field in a pancake-shaped BEC. The MW field
is polarized in the direction perpendicular to the pancake's plane.}
\label{fig1}
\end{figure}

\section{The model}

As schematically shown in Fig. \ref{fig1}, we consider a nearly-2D
(pancake-shaped) binary BEC composed of two different hyperfine states of
the same atomic species, which is described by the two-component
(pseudo-spinor) wave function, $\left\vert \Psi \right\rangle =\left( \Psi
_{\downarrow },\Psi _{\uparrow }\right) ^{T}$, with each component emulating
``spin-up" and ``spin-down" states. The
corresponding Hamiltonian is $\mathcal{H}=\hat{\mathbf{p}}^{2}/2m_{\mathrm{at%
}}-\left( \hbar \delta /2\right) \sigma _{3}-\mathbf{m}\cdot \mathbf{B}$
\cite{dong3}, where $m_{\mathrm{at}}$, $\hat{\mathbf{p}}$, and $\mathbf{m}$\
are the atomic mass, 2D momentum, and magnetic moment, $\hbar \delta $\ an
energy difference between atomic states $\left\vert \uparrow \right\rangle $%
\ and $\left\vert \downarrow \right\rangle $, $\sigma _{3}$\ the Pauli
matrix, and $\mathbf{B}=\mu _{0}(\mathbf{H}+\mathbf{M)}$\ is the magnetic
induction, with magnetic field $\mathbf{H}$\ and magnetization $\mathbf{M}%
=\left\langle \Psi \right\vert \mathbf{m}\left\vert \Psi \right\rangle $. In
the rotating-wave approximation, the atomic wave function, $\left\vert \psi
\right\rangle \equiv \left\vert \phi \right\rangle e^{\pm i\omega t/2}\equiv
\left( \phi _{\downarrow },\phi _{\uparrow }\right) ^{T}$, is governed by
coupled Gross-Pitaevskii equations (GPEs), with $\ast $ standing for the
complex conjugate:
\begin{eqnarray}
i\hbar \partial \phi _{\downarrow }/\partial t &=&\left( \hat{\mathbf{p}}%
^{2}/2m_{\mathrm{at}}+\hbar \Delta /2\mathbf{-}\mu _{0}\mathbf{\mathbf{%
\mathbf{m}_{\uparrow \downarrow }\cdot }\mathbf{m}_{\downarrow \uparrow }}%
\left\vert \phi _{\uparrow }\right\vert ^{2}\right) \phi _{\downarrow }\notag\\
&&-\mu_{0}\mathbf{m}_{\downarrow \uparrow }\cdot \mathbf{H}^{\ast }\phi _{\uparrow},  \notag \\
i\hbar \partial \phi _{\uparrow }/\partial t &=&\left( \hat{\mathbf{p}}%
^{2}/2m_{\mathrm{at}}-\hbar \Delta /2\mathbf{-\mu _{0}\mathbf{\mathbf{m}%
_{\uparrow \downarrow }\cdot }\mathbf{m}_{\downarrow \uparrow }}\left\vert
\phi _{\downarrow }\right\vert ^{2}\right) \phi _{\uparrow }\notag\\
&&-\mu _{0}\mathbf{m}_{\uparrow \downarrow }\cdot \mathbf{H}\phi _{\downarrow },  \label{GPE}
\end{eqnarray}
with detuning $\Delta =\omega -\delta $ of the MW from the atomic
transition, and matrix elements of the magnetic moment, $\mathbf{m}%
_{\uparrow \downarrow }$ and $\mathbf{m}_{\downarrow \uparrow }$ ($\mathbf{m}%
_{\uparrow \uparrow }=\mathbf{m}_{\downarrow \downarrow }=0$ due to the
symmetry).

The magnetic field and magnetization, which are polarized perpendicular to
the pancake's plane, are each represented by a single component, $H$ and $M$%
, which obey the Helmholtz equation,
\begin{equation}
\nabla ^{2}H+k^{2}H=-k^{2}M,  \label{Helmholtz}
\end{equation}%
\ where $k$\ is the MW wavenumber. As the wavelength of the MW field, $%
\lambda =2\pi /k$, is always much greater than an experimentally relevant
size of the BEC, the second term in Eq. (\ref{Helmholtz}) may be omitted in
comparison with the first term (see also Ref. \cite{dong3}), reducing Eq. (%
\ref{Helmholtz}) to the Poisson equation for the scalar field:
\begin{equation}
\nabla ^{2}H=-k^{2}M.  \label{Poisson}
\end{equation}%
Because the medium's magnetization, which is the source of the magnetic
field, is concentrated in the pancake, the Poisson equation may be treated
as one in the 2D plane. Them using the Green's function of the 2D Poisson
equation, the magnetic field is given by\textbf{\ }%
\begin{equation}
H=H_{0}-Nk^{2}\left\vert \mathbf{m_{\downarrow \uparrow }}\right\vert /(2\pi
l_{\perp })\int \!\!\ln \left( \left\vert \mathbf{r}-\mathbf{r^{\prime }}%
\right\vert \right) \phi _{\downarrow }^{\ast }\left( \mathbf{r^{\prime }}%
\right) \phi _{\uparrow }\left( \mathbf{r^{\prime }}\right) d\mathbf{r}%
^{\prime }\mathbf{,}  \label{H}
\end{equation}%
\textbf{\ }where $H_{0}$\ is a background magnetic field of the MW, $N$ is
the number of atoms, and $\mathbf{r}$\ is the set of 2D coordinates
normalized by the transverse confinement size $l_{\perp }$. Then, the GPEs
for the wave function, subject \ to normalization $\int \left( \left\vert
\phi _{\uparrow }\right\vert ^{2}+\left\vert \phi _{\downarrow }\right\vert
^{2}\right) d\mathbf{r}=1$, takes the form of\textbf{\ }%
\begin{gather}
i\frac{\partial \phi _{\downarrow }}{\partial \tau }=\left( -\frac{1}{2}%
\nabla ^{2}+\eta +H_{0}-\beta \left\vert \phi _{\uparrow }\right\vert
^{2}\right) \phi _{\downarrow }  \notag \\
+\frac{\gamma \phi _{\uparrow }}{2\pi }\int \!\!\ln \left( \left\vert
\mathbf{r}-\mathbf{r^{\prime }}\right\vert \right) \phi _{\downarrow }\left(
\mathbf{r^{\prime }}\right) \phi _{\uparrow }^{\ast }\left( \mathbf{%
r^{\prime }}\right) \!d\mathbf{r^{\prime }},  \label{eq:final-down}
\end{gather}%
\begin{gather}
i\frac{\partial \phi _{\uparrow }}{\partial \tau }=\left( -\frac{1}{2}\nabla
^{2}-\eta +H_{0}-\beta \left\vert \phi _{\downarrow }\right\vert ^{2}\right)
\phi _{\uparrow }  \notag \\
+\frac{\gamma \phi _{\downarrow }}{2\pi }\int \!\!\ln \left( \left\vert
\mathbf{r}-\mathbf{r^{\prime }}\right\vert \right) \phi _{\downarrow }^{\ast
}\left( \mathbf{r^{\prime }}\right) \phi _{\uparrow }\left( \mathbf{%
r^{\prime }}\right) \!d\mathbf{r^{\prime }},  \label{eq:final-up}
\end{gather}%
where rescaling is defined by $\phi _{\uparrow ,\downarrow }=\sqrt{N}%
/l_{\perp }\psi _{\uparrow ,\downarrow }$, $\tau =t/t_{0}$\ with $%
t_{0}=\hbar /E_{c}$\ and\textbf{\ }$E_{c}=\hbar ^{2}/(m_{\mathrm{at}%
}l_{\perp }^{2})$\textbf{, }$\eta \equiv t_{0}\Delta /2$, and scaled
strengths of the LFE and contact interactions (if any) are \textbf{\ }%
\begin{equation}
\gamma =m_{\mathrm{at}}l_{\perp }k^{2}N\mu _{0}\left\vert \mathbf{%
m_{\downarrow \uparrow }}\right\vert ^{2}/\hbar ^{2}\mathbf{,\ }\beta \equiv
N\mu _{0}\mathbf{m}_{\uparrow \downarrow }\cdot \mathbf{m}_{\downarrow
\uparrow }/\left( \hbar l_{\perp }^{3}E_{c}\right) \mathbf{.}  \label{gamma}
\end{equation}%

To describe experimental conditions, three-dimensional should also include
the trapping potential, $\left( \Omega ^{2}/2\right) r^{2}\phi _{\uparrow
,\downarrow }$. It has been checked that, after the creation of the trapped
modes, the potential may be switched off, leading to smooth transformation
of the modes into their self-trapped counterparts obtained directly found in
the free space ($\Omega =0$). The vorticity may be imparted to the trapped
condensate by a vortical optical beam \cite{storage-vortex}.

If collisions between atoms belonging to the two components are considered
(with the corresponding strength of the contact interaction tunable by dint
of the Feshbach resonance \cite{FR}), the additional cross-cubic terms can
be absorbed into rescaled coefficient $\beta $. Collisions may also give
rise to self-interaction terms, $-\tilde{\beta}\left\vert \phi _{\downarrow
}\right\vert ^{2}$ and $\tilde{\beta}\left\vert \phi _{\uparrow }\right\vert
^{2}$, in the parentheses of Eqs. (\ref{eq:final-down}) and (\ref%
{eq:final-up}), respectively. On the other hand, the same equations with $%
\tilde{\beta}=0$ apply as well to a different physical setting, \textit{viz}%
., a degenerate Fermi gas with spin $1/2$, in which $\phi _{\downarrow }$
and $\phi _{\uparrow }$ represent two spin components, coupled by the MW
magnetic field \cite{Fermi,dong3}.

\begin{figure}[tbp]
\centering
\includegraphics{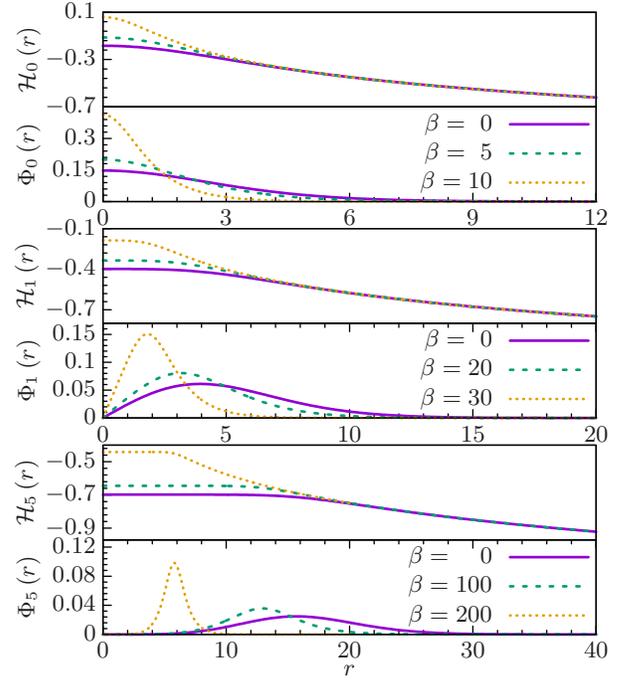}
\caption{Radial profiles of the magnetic filed and wave functions in
fundamental solitons (top) and vortices with $S=1$ (middle) and $S=5$
(bottom) at indicated values of $\protect\beta $.}
\label{fig2}
\end{figure}

The following analysis is chiefly dealing with the zero-detuning (symmetric)
system, $\eta =0$. In this case, Eqs. (\ref{eq:final-down}) and (\ref%
{eq:final-up}) coalesce into a single equation for $\phi _{\downarrow }=\phi
_{\uparrow }\equiv \phi \exp \left( -iH_{0}\tau\right) $, subject to
normalization $\int \left\vert \phi (\mathbf{r})\right\vert ^{2}d\mathbf{r}%
=1/2$:%
\begin{equation}
i\frac{\partial \phi }{\partial \tau}=\left[ -\frac{1}{2}\nabla ^{2}-\beta
\left\vert \phi \right\vert ^{2}+\frac{\gamma }{2\pi }\int \!\!\ln \left(
\left\vert \mathbf{r}-\mathbf{r^{\prime }}\right\vert \right) \left\vert
\phi \left( \mathbf{r^{\prime }}\right) \right\vert ^{2}\!d\mathbf{r^{\prime
}}\right] \phi ,  \label{phi}
\end{equation}%
and the above-mentioned self-interaction coefficient, $\tilde{\beta}$, may
be absorbed into $\beta $. This equation and the normalization condition are
invariant with respect to the self-similarity transformation: $\phi \left(
\mathbf{r},\tau\right) =\sqrt{\gamma _{0}}\tilde{\phi}\left( \mathbf{\tilde{r}}%
,\tau\right) \exp \left\{ -i\left[ \gamma \left( \ln \gamma _{0}\right) /8\pi %
\right] \tau\right\} ,$ $\tau=\gamma _{0}^{-1}\tilde{\tau},$ $\mathbf{r}=\gamma
_{0}^{-1/2}\mathbf{\tilde{r}},$ $\gamma =\gamma _{0}\tilde{\gamma}$, $\beta
\equiv \tilde{\beta}$ , which allows one to replace $\gamma $ by $\gamma
/\gamma _{0}$ with arbitrary factor $\gamma _{0}$. We use this option to to
fix $\gamma =\pi $ in the numerical analysis of the symmetric configuration.
In physical units, for alkali atoms transversely confined with $l_{\perp }=1~%
\mathrm{\mu }$m and irradiated by a MW with wavelength $1$\ mm, the above
definition yields $\gamma \sim 10^{-7}N$. Thus, $\gamma \sim 10$ for the
experimentally available ``massive" BEC with $N\sim 10^{8}$\
\cite{large,large1}, while a typical VA radius can be estimated as $%
\sim 10$\ $\mathrm{\mu }$m (see Figs. \ref{fig2} and \ref{fig5} below), and
a characteristic range of the magnetic-field amplitudes may reach a few
gauss.

\section{The results}

Stationary solutions to Eq. (\ref{phi}) with chemical potential $\mu $ and
vorticity $S$ are looked for, in polar coordinates $\left( r,\theta \right) $%
, as
\begin{equation}
\phi =e^{-i\mu \tau -iS\theta }\Phi _{S}\left( r\right) \mathbf{,}
\label{Phi}
\end{equation}%
\textbf{\ }where $\Phi _{S}(r)$ is a real radial wave function. Typical
examples of solutions for $\Phi _{S}\left( r\right) $, produced by the
imaginary-time evolution method \cite{IT}, are plotted in Fig. (\ref{fig2}),
for different values of $S$\ and $\beta \geq 0$. 
Numerical results demonstrate that fundamental solitons (which correspond to
$S=0$) and VAs are destroyed by the collapse at $\beta >\beta _{\max }(S)$,
see Table \ref{table1}. This critical value can be found by considering the
energy corresponding to Eqs. (\ref{eq:final-down}) and (\ref{eq:final-up})
with $\phi _{\uparrow }=\phi _{\downarrow }$,%
\begin{gather}
E=2\pi \int_{0}^{\infty }rdr\left[ (\Phi _{S}^{\prime })^{2}+r^{-2}S^{2}\Phi
_{S}^{2}-\beta \Phi _{S}^{4}\right]   \notag \\
+\frac{\gamma }{2\pi }\int \int d\mathbf{r}_{1}d\mathbf{r}_{2}\ln \left(
\left\vert \mathbf{r}_{1}-\mathbf{r}_{2}\right\vert \right) \Phi _{S}^{2}(%
\mathbf{r}_{1})\Phi _{S}^{2}(\mathbf{r}_{2}).  \label{E}
\end{gather}%
The numerical findings displayed in Figs. \ref{fig2} and \ref{fig5} suggest
that, for $S\geq 2$ and $\beta $ large enough, the vortex takes the shape of
a narrow annulus, which may be approximated by the usual quasi-1D soliton
shape in the radial direction, with regard to the adopted normalization, cf.
Ref. \cite{Caplan}:
\begin{equation}
\Phi _{S}(r)=\sqrt{\beta }/\left( 8\pi R\right) \mathrm{sech}[\beta \left(
r-R\right) /\left( 8\pi R\right) ],  \label{narrow}
\end{equation}%
where $R$ is the VA's radius. The substitution of this approximation in Eq. (%
\ref{E}) yields%
\begin{equation}
E(R)=\left[ S^{2}-\frac{\beta ^{2}}{3\left( 8\pi \right) ^{2}}\right] \frac{1%
}{2R^{2}}+\frac{\gamma }{8\pi }\ln R.  \label{Eeff}
\end{equation}%
Next, the annulus' radius $R$ is to be selected as a point corresponding to
the energy minimum: $dE/dR=0$, i.e., $R_{\min }^{2}=(8\pi /\gamma )\left[
S^{2}-(1/3)\left( \beta /8\pi \right) ^{2}\right] $ (comparison with
numerical results demonstrates that $R_{\min }$ provides a reasonable
approximation for the radius of narrow VAs). Then, $\beta _{\max }$ is
predicted as the value at which $R_{\min }^{2}$ vanishes, i.e., the annulus
collapses to the center,
\begin{equation}
\beta _{\max }^{\mathrm{(an)}}=8\sqrt{3}\pi S.  \label{beta_max}
\end{equation}%
As seen in Table \ref{table1}, this analytical prediction is virtually
identical to its numerically found counterparts at $S\geq 2$.

Further, it is found that $\beta _{\max }$ is the same as in the
``simplified" 2D GPE that contains solely the
local-attraction term,
\begin{equation}
i\partial \phi /\partial \tau =-\left[ (1/2)\nabla ^{2}+\beta \left\vert
\phi \right\vert ^{2}\right] \phi ,  \label{simple}
\end{equation}
for which the existence limit was found in Refs. \cite{Townes}, for $S=0$,
and in Ref. \cite{Minsk} for $1\leq S\leq 5$ , i.e., $\beta _{\max }$ does
not depend of the LFE strength, $\gamma $.
To explain this fact, we note that, at the limit stage of the collapse, when
the shrinking 2D annulus becomes extremely narrow, the equation for the wave
function becomes asymptotically tantamount to Eq. (\ref{simple}), therefore
the condition for the onset of the collapse is identical in both equations.
However, the solitons of Eq. (\ref{simple}) exist solely at $\beta =\beta
_{\max }$, being completely unstable, while the LFE-induced long-range
interaction in Eqs. (\ref{eq:final-down}) and (\ref{eq:final-up}) creates
\emph{stable} solitons and vortices for all $S$, as shown below. It is
worthy to stress too that the analytical result given by Eq. (\ref{beta_max}%
) provides an explanation for the numerical findings that were first
reported in Ref. \cite{Minsk} and later considered in many works, but never
reproduced in an analytical form.

\begin{table}[tbp] \centering%
\begin{tabular}{|l|l|l|l|l|l|l|l|}
\hline
$S$ & $\beta _{\max }$ & $\beta _{\max }^{\mathrm{(an)}}$ & $\beta _{\mathrm{%
st}}$ & $S$ & $\beta _{\max }$ & $\beta _{\max }^{\mathrm{(an)}}$ & $\beta _{%
\mathrm{st}}$ \\ \hline
$0$ & $11.8$ & $\mathrm{n/a}$ & $\equiv \beta _{\max }$ & $3$ & $132.5$ & $%
130.6$ & $41$ \\ \hline
$1$ & $48.3$ & $43.5$ & $11$ & $4$ & $175.5$ & $174.1$ & $57$ \\ \hline
$2$ & $89.7$ & $87.0$ & $28$ & $5$ & $218.5$ & $217.7$ & $70$ \\ \hline
\end{tabular}
\caption{$\beta _{\max }$ and $\beta _{\max }^{\mathrm{(an)}}$: numerically obtained
and analytically predicted values of the contact-interaction strength, $\beta$,
up to which the fundamental solitons and vortex annuli exist.
$\beta _{\mathrm{st}}$: the numerically identified
stability boundary of the vortex annuli.}\label{table1}%
\end{table}%

\begin{figure}[tbp]
\centering
\includegraphics{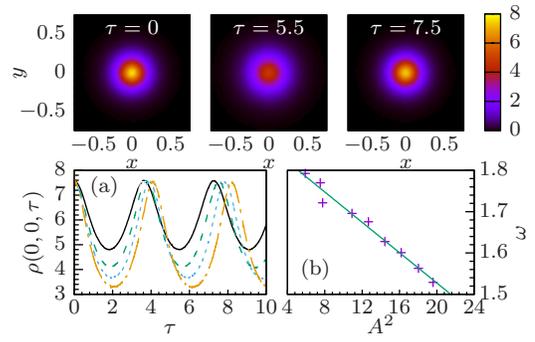}
\caption{(a) Oscillations of the peak density, $\protect\rho \left(
0,0\right) \equiv \left\vert \protect\phi _{\downarrow }(x=y=0)\right\vert
^{2}$ $+\left\vert \protect\phi _{\uparrow }\left( x=y=0\right) \right\vert
^{2}$, of the perturbed fundamental soliton at $\protect\beta =11.6$ (for
different perturbation amplitudes). (b) The oscillation frequency vs. the
squared oscillation amplitude, $A^{2}$. The top row displays profiles of the
oscillating soliton.}
\label{fig3}
\end{figure}

The stability of the self-trapped modes has been systematically tested by
real-time simulations of Eqs. (\ref{eq:final-down}) and (\ref{eq:final-up})
with random perturbations added to the stationary solutions (independent
perturbations were taken for $\phi _{\uparrow }$\ and $\phi _{\uparrow }$,
to verify the stability against breaking the symmetry between them). The
fundamental solitons are stable in their entire existence region, $\beta
<\beta _{\max }\approx 11.8$. At $\beta $\ very close to $\beta _{\max }$,
the perturbations lead to persistent oscillations, as shown in Fig. \ref%
{fig3}(a) for $\beta =11.6$, due to excitation of a soliton's internal mode
\cite{IM,IM1,IM2}. It is seen in Fig. \ref{fig3}(b) that the oscillation
frequency is a nearly linear function of the squared amplitude of the
oscillations, which is a typical feature of a nonlinear oscillatory mode.


Systematic simulations of the evolution of the VA families reveal an
internal stability boundary, $\beta _{\mathrm{st}}(S)<\beta _{\max }(S)$
(see Table 1), the vortices being stable at $\beta <\beta _{\mathrm{st}}(S)$%
. In the interval of $\beta _{\mathrm{st}}(S)<\beta <\beta _{\max }(0)$,
they are broken by azimuthal perturbations into rotating necklace-shaped
sets of fragments, which resembles the initial stage of the instability
development of localized vortices in usual models \cite%
{trap,non-Kerr,split1,split}; however, unlike those models, the necklace
does not expand, remaining confined under the action of the effective
nonlocal interaction. Typical examples of the stable and unstable evolution
of fundamental solitons and VAs are displayed, respectively, in Figs. \ref%
{fig4} and \ref{fig5}.

\begin{figure}[tbp]
\centering\includegraphics{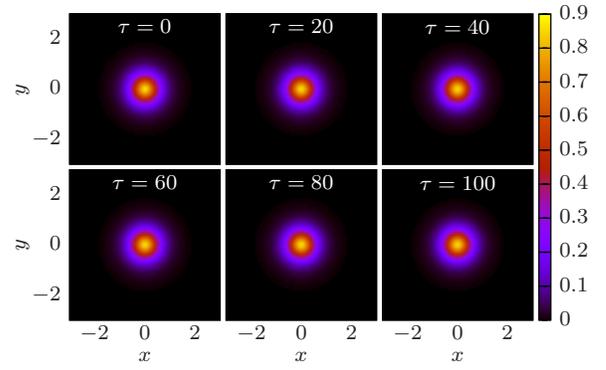}
\caption{Stable perturbed evolution of the fundamental soliton, with $S=0$ and $%
\protect\beta =11$. Note that this value of $\protect\beta $ is close to the
existence boundary, $\protect\beta _{\max }(S=0)=11.8$, see Table 1.}
\label{fig4}
\end{figure}

\begin{figure}[tbp]
\centering
\includegraphics{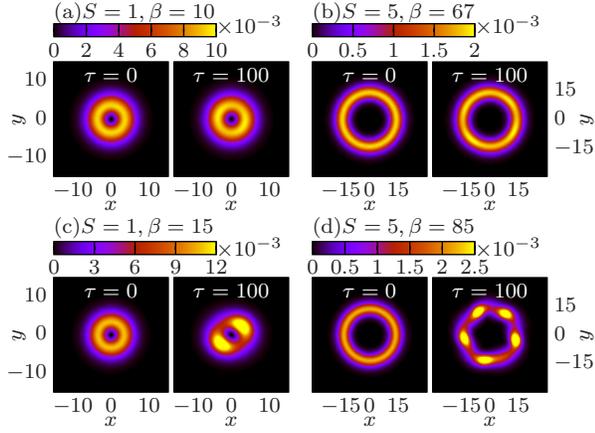}
\caption{Top and bottom panels display, severally, examples of the stable
and unstable perturbed evolution of the VAs with indicated values of $S$ and $\protect%
\beta $. The necklace-shaped set, observed in the latter case, remains
confined (keeping the same overall radius) in the course of subsequent
evolution.}
\label{fig5}
\end{figure}


To address the stability of the VAs against azimuthal perturbations in an
analytical form, we approximate the wave function of a perturbed VA by $%
A(\theta )\Phi _{S}(r)$ and derive an evolution equation for the modulation
amplitude, $A$, by averaging Eqs. (\ref{eq:final-down})-(\ref{eq:final-up})
in the radial direction:
\begin{equation}
i\frac{\partial A}{\partial \tau }=-\frac{1}{2R^{2}}\frac{\partial ^{2}A}{%
\partial \theta ^{2}}+\left[ \frac{\gamma \ln R}{4\pi R}-\frac{2\beta ^{2}}{%
3\left( 8\pi R\right) ^{2}}\right] |A|^{2}A.  \label{NLS}
\end{equation}%
A straightforward analysis of the modulational stability of the solution
with $A=1$ against perturbations $\sim \exp \left( ip\theta \right) $ with
integer winding numbers $p$ \cite{KA} shows that the stability is maintained
under the threshold condition, $p^{2}\geq \left( 8/3\right) \left( \beta
/8\pi \right) ^{2}$, if the term $\sim \beta ^{2}$ dominates in Eq. (\ref%
{NLS}). Further, the numerical results demonstrate that, as in other models
\cite{other}, the critical instability corresponds to $p^{2}=S^{2}$ (for
instance, the appearance of five fragments in the part of Fig. \ref{fig5}
corresponding to $S=5,\beta =85$ demonstrates that, for $S=5$, the dominant
splitting mode has $p=5$). Thus, it is expected that the VAs remain stable
at $\beta <\beta _{\mathrm{st}}^{(\mathrm{an})}(S)=2\sqrt{6}\pi S\approx
\allowbreak 15.4S$. On the other hand, the numerically found stability
limits collected in Table 1 obey an empirical formula, $\beta _{\mathrm{st}%
}^{(\mathrm{num})}(S)\approx \allowbreak 15S-4$. Thus, the analytical
approximation is quite accurate for $S\geq 2$. To put this result in a
physical context, we note that, in terms of experimentally relevant
parameters, the scaling adopted above implies $|\beta |\sim \left(
|a_{s}|/l_{\perp }\right) N$, where $a_{s}<0$ is the scattering length which
accounts for the contact attraction. Thus, values of $\beta $ (actually, of
either sign) may be relevant up to $|\beta |\sim 1000$.
\begin{figure}[tbp]
\centering
\includegraphics{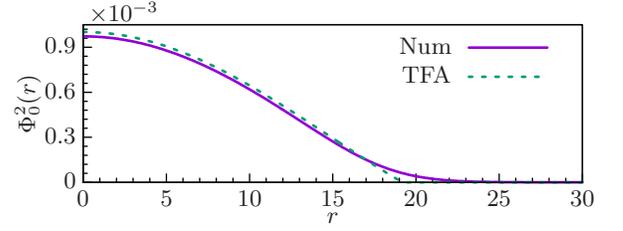}
\caption{Comparison of the Thomas-Fermi approximation, as given by Eq. (%
\protect\ref{TF}), for the fundamental soliton (the dashed line) and its
numerically found counterpart (the solid line), for $\protect\beta =-200$.}
\label{fig6}
\end{figure}


It follows from these results that the giant VAs, with higher values of $S$,
are \emph{much more robust} than their counterparts with smaller $S$. This
feature is opposite to what was previously found in those (few) models which
are able to produce stable VAs with $S>1$ \cite{non-Kerr,Barcelona,Nal}. It
is relevant to mention that, at $\beta <\beta _{\mathrm{st}}(S=0)$, the
fundamental soliton is the system's ground state, while, at $\beta $ $>$ $%
\beta _{\mathrm{st}}(S=0)$, the ground state does not exist, due to the
possibility of the collapse. The vortices with $\beta _{\mathrm{st}%
}(S)>\beta $ cannot represent the ground state, but, nevertheless, they
exist as metastable ones, cf. the spin-orbit-coupled system, considered in
Ref. \cite{SOC3D}, where self-trapped three-dimensional modes of the
semi-vortex type exist too as metastable states, although the system does
not have a ground state, due to the presence of the supercritical collapse.

For the strong \emph{repulsive} contact interaction (large $\beta <0$),
fundamental solitons (with $S=0$) can be constructed by means of the
Thomas-Fermi approximation (TFA), as shown by straightforward consideration
of the stationary version of Eq. (\ref{phi}), with the substitution of the
stationary wave form as per Eq. (\ref{Phi}). In this case, it is more
convenient, instead of using the Green's function, to explicitly combine the
stationary equation with Poisson equation (\ref{Poisson}). The result is
\begin{equation}
\left( \Phi _{0}^{2}\right) _{\mathrm{TFA}}\left( r\right) =\left\{
\begin{array}{c}
\phi _{0}^{2}J_{0}\left( \xi r\right) ~~\mathrm{at~~}r<r_{1}/\xi ,\  \\
\left( \Phi _{0}^{2}\right) _{\mathrm{TFA}}\left( r\right) =0~~\mathrm{at}%
~~r>r_{1}/\xi \ ,%
\end{array}%
\right.  \label{TF}
\end{equation}
where $\xi \equiv \sqrt{\gamma /\left\vert \beta \right\vert }$, $%
r_{1}\approx 2.4$\ is the first zero of Bessel function $J_{0}\left(
r\right) $, and $\phi _{0}$\ is a normalization constant. Figure \ref{fig6}
shows that the TFA agrees very well with the numerical solution.

Lastly, it is relevant to proceed from the symmetric system [$\eta =0$, $%
\phi _{\uparrow }=\phi _{\downarrow }$ in Eqs. (\ref{eq:final-down}) and (%
\ref{eq:final-up})] to a strongly asymmetric one, with large $\eta $. The
relevant solution has $\mu =-\eta +\delta \mu $ with $\left\vert \delta \mu
\right\vert \ll \eta $\textbf{\ }and small component\textbf{\ }$\Phi
_{\downarrow }\approx \left( H_{0}/2\eta \right) \Phi _{\uparrow }$, while
the large one satisfies equation
\begin{eqnarray}
\left( \Delta \mu +\frac{H_{0}^{2}}{2\eta }\right) \Phi _{\uparrow }&=-\frac{1%
}{2}\nabla ^{2}\Phi _{\uparrow }-\frac{\beta H_{0}^{2}}{4\eta ^{2}}\Phi
_{\uparrow }^{3}-\frac{\gamma H_{0}^{2}}{8\pi \eta ^{2}}\Phi _{\uparrow
}\notag\\
&\int \ln \left( \left\vert \mathbf{r}-\mathbf{r}^{\prime }\right\vert
\right) \Phi _{\uparrow }^{2}(\mathbf{r}^{\prime })d\mathbf{r}^{\prime }.
\label{asymm}
\end{eqnarray}%
Up to obvious rescaling, Eq. (\ref{asymm}) is identical to the equation for
the stationary\ wave function in the symmetric case, i.e., Eq. (\ref{phi})
with substitution of the wave function as per Eq. (\ref{Phi}), with any
value of $S$. Thus, the strongly asymmetric solutions can obtained by means
of the rescaling of their symmetric counterparts.

\section{Conclusion}

In this work we have developed the analysis for the 2D fundamental solitons
and VAs (vortex annuli) produced by the LFE (local field effect) in the BEC
composed of two-level atoms, or, alternatively, a gas of fermions, in which
two components are coupled by the MW (microwave) field. The effective
long-range interaction mediated by the field stabilizes the solitons and
VAs, even in the presence of the attractive contact interaction between the
two components, which, by itself, leads to the critical collapse. The
solitons exists too and are stable in the presence of the arbitrarily strong
contact repulsion. Nearly exact critical values of the local-attraction
strength, $\beta _{\max }$, up to which the solitons and vortices exist,
have been found analytically. This result also provides an analytical
explanation to the well-known existence limits of VAs in the 2D nonlinear
Schr\"{o}dinger/Gross-Pitaevskii equation with the cubic self-focusing term,
which were previously known solely in the numerical form. While the
fundamental solitons are stable up to $\beta =\beta _{\max }$, the VAs
remain stable in a smaller interval, $\beta \leq \beta _{\mathrm{st}}<\beta
_{\max }$, being vulnerable to the azimuthal instability at $\beta _{\mathrm{%
st}}<\beta <\beta _{\max }$. The stability boundary, $\beta _{\mathrm{st}}$,
is found in an approximate analytical form too. On the contrary to
previously studied models \cite{non-Kerr,Barcelona,Nal}, the (giant)\textbf{%
\ }VAs with higher vorticities, such as $S=5$, are more robust than their
counterparts with small $S$. In addition, a very accurate TFA (Thomas-Fermi
approximation) was developed for the fundamental solitons, with $S=0$. The
results have been obtained for both symmetric and strongly asymmetric
two-component systems.

The VAs obtained here can be further used to construct vortex lattices \cite%
{zeng}. Challenging possibilities are to consider interaction between the
self-trapped modes, and, eventually, to extend the model to the fully 3D
setting. Another direction for the extension of the work is to explore the
electric LFE in a molecular condensate.

\section*{Acknowledgments}

This work was supported, in part, by National Science Foundation of China
(grants Nos. 11574085 and 91536218), the Research Fund for the Doctoral
Program of Higher Education of China (grant No. 20120076110010), Program of
Introducing Talents of Discipline to Universities (B12024), and by the joint
program in physics between the National Science Foundation (US) and
Binational Science Foundation (US-Israel), through grant No. 2015616. B.A.M.
appreciates hospitality of the State Key Laboratory of Precision
Spectroscopy and Department of Physics at the East China Normal University.

\end{document}